
\input amstex
\documentstyle{amsppt}
\magnification=1200
\topmatter
\title
Ricci Fall-off in Static, Globally Hyperbolic,\\
Non-singular Spacetimes
\endtitle
\rightheadtext{Ricci Fall-off in Non-singular Spacetimes}
\author
David Garfinkle \\  Steven G. Harris
\endauthor
\address
Department of Physics, Oakland University,
Rochester, MI 48309 \endaddress
\email
garfinkl\@oakland.edu \endemail
\address
Department of Mathematics, St. Louis University,
St. Louis, Mo 63103, USA\endaddress
\email harrissg\@sluvca.slu.edu\endemail

\abstract
What restrictions are there on a spacetime for which the
Ricci curvature is such as to produce convergence of
geodesics (such as the preconditions for the Singularity
Theorems) but for which there are no singularities? We
answer this question for a restricted class of spacetimes:
static, geodesically complete, and globally hyperbolic.
The answer is that, in at least one spacelike direction,
the Ricci curvature must fall off at a rate inversely
quadratic in a  naturally-occurring Riemannian metric on
the space of static observers.  Along the way, we establish
some global results on the static observer space, regarding
its completeness and its behavior with respect to universal
covering spaces.
\endabstract

\endtopmatter
\document
\head
1. Introduction
\endhead
The Ricci curvature of a spacetime is what is used to
drive the  Singularity Theorems of Hawking and Penrose
\cite{1,2}.  Essentially, if $Ric$ along a geodesic has a
positive lower bound, then a conjugate  point must occur on
the geodesic within a certain length \cite{3,4}. However,
in a globally hyperbolic spacetime timelike-related points
must be joined by maximal geodesics (no conjugate points).
One can then obtain a contradiction if Ricci curvature is
sufficiently large and all geodesics are complete.

So suppose we disallow singularities---assume geodesic
completeness---but keep global hyperbolicity:  What is
forced  to happen?  Just how must the Ricci curvature
behave?

To simplify matters, we shall also assume a static
spacetime.  The energy condition we shall use is the null
energy condition: $Ric(N,N) \ge 0$ for all null vectors
$N$.

In section 2 we prove some results about the space of
static observers.  Section 3 contains our result on the
behavior of the Ricci tensor.  In section 4 we discuss the
nature of this result and consider two examples of
the behavior of the Ricci tensor in spacetimes satisfying
our conditions.
\head
2. The space of static observers
\endhead
Our spacetime $M$ comes equipped with a field of
distinguished  observers,  the static observers (i.e., the
integral curves of the timelike Killing field); this
amounts to a one-dimensional foliation of
$M$.  Any foliation $\Cal{F}$ of a manifold $M$ gives
rise to the leaf space $Q$, the quotient of $M$ by the
equivalence relation of two points being equivalent if they
lie on the same leaf  of the foliation.  For us, $\Cal{F}$
is the set of static observer worldlines; the space $Q$ of
static observers is the most natural object to use for
analyzing the geometry of $M$.

Geroch \cite{5}, in the context of a stationary
spacetime, introduced the notion of examining the observer
space $Q$.  In the general stationary spacetime, there is
no guarantee that $Q$ is even remotely well-behaved
topologically (it can be totally indiscrete!).  However, it
is shown in Harris \cite{6} that a chronological stationary
spacetime must have an observer space that is almost a
manifold: locally Euclidean, but possibly not Hausdorff.

In order to show that $Q$ is fully a manifold (i.e., that
it is Hausdorff), all we need do is show that the Killing
field $U$ is a complete vector field \cite{6}.  It is shown
in Proposition 9.30 of O'Neill \cite{7} that geodesic
completeness implies completeness of the Killing field.
However, we want to use the  weaker condition of either
timelike or null geodesic completeness.  We now show

\proclaim{Lemma 1}
Let ($M,g$) be a stationary spacetime that is timelike or
null  geodesically complete.  Then the Killing field is
complete.
\endproclaim

\demo{Proof}
For any point $p$ let $\gamma_p$ be the integral curve of
the Killing field $U$ for which ${\gamma_p}(0) = p$.  Let
$\left({a_p},{b_p}\right)$ be the maximum interval on which
this curve is defined.  Let $q$ be a point not on
$\gamma_p$ such that there is a timelike (respectively,
null) geodesic from $p$ to $q$ and also from
$q$ to a point $p'\ne p$ on $\gamma_p$.  We will show
first that $a_q \le a_p$ and $b_q \ge b_p$.

We proceed by constructing a two-surface $\Sigma$ with
boundary $\gamma_p$ and $\gamma_q$ and ruled by timelike
(respectively, null) geodesics.  Let $\zeta_0$ be the
geodesic from $p$ to $q$ with ${\zeta_0}(0) = p$ and
${\zeta_0}(1) = q$.  Let $X$ be the tangent vector to
$\zeta_0$ at $p$ and extend $X$ along $\gamma_p$ by $[X,U]
= 0$; note that since $U$ is Killing, its flow is an
isometry, so $X$ is timelike (respectively, null) at all
points.  Now let $\zeta_t$ be the geodesic with
${\zeta_t}(0) = {\gamma_p}(t)$ and ${{\zeta_t}'}(0) = X$ (by
timelike or null geodesic completeness, ${\zeta_t}(s) $ is
defined for all $s$); let $\Sigma = \{\zeta_t(s) | t \in (a_p,
b_p), s \in [0,1]\}$.  Note that $\zeta_ t$ is the image of
$\zeta_0$ under the isometry generated by $U$ within
$\Sigma$; in particular, ${\zeta_t}(1) = {\gamma_q}(t)$.
In other words, $\gamma_q$ is defined for at least the same
parameter values as is $\gamma_p$: $a_q \le a_p$ and $b_q
\ge b_p$.

By the same argument, $a_{p'} \le a_q$ and $b_{p'} \ge
b_q$: $\left(a_{p'},b_{p'}\right) \supset
\left(a_p,b_p\right)$; but, since $p$ and $p'$ are on the
same integral curve of $U$, this can happen only if $a_p =
-\infty$ and $b_p = \infty$.  It follows that $U$ is
complete.
\qed\enddemo

Now using the results of \cite{6} we have

\proclaim{Theorem 2}
Let $M$ be a timelike or null geodesically complete,
chronological, stationary spacetime.  Then the space $Q$ of
stationary observers is a (Hausdorff) manifold.
\qed\endproclaim

The observer space $Q$ tells quite a lot about the
topology and causal structure of $M$:  It contains all of
the information on the ``spacelike topology" of $M$.  This
is so in a rather strong sense:

\proclaim{Theorem 3}
Let $M$ and $Q$ be as in the previous theorem.  Then $M$ is
diffeomorphic to $Q\times\Bbb R$.  Furthermore, for any
edgeless,  achronal, embedded spacelike hypersurface $N$ in
$M$, $N$ must be diffeomorphic to $Q$.
\endproclaim

\demo{Proof}  Let $\pi: M \to Q$ be the projection to the
orbit space,  making $M$ a smooth line-bundle over $Q$ (the
differentiable structure on $Q$ comes from local
identification with a cross-section in a flow-box for the
Killing field).  Since the basespace $Q$ is a (Hausdorff)
manifold, the bundle has global cross-sections, and $M$ is
diffeomorphic to $Q\times\Bbb R$ (see, for instance,
\cite{10}, Theorem I.5.7).  Let $N$ be a spacelike
hypersurface in $M$.  We will see that $\pi$, restricted to
$N$, provides a diffeomorphism with $Q$, under the right
circumstances.

Since $N$ is spacelike and of the same dimension as $Q$
and is a manifold without boundary, and $\pi_*$ kills only
timelike vectors, $\pi(N)$ must be an open subset of $Q$.
With $N$ achronal, $\pi$ must be injective on $N$ ($\pi(x) =
\pi(y)$ implies there is a Killing orbit from $x$ to $y$,
which is a timelike curve between the two points).  It
follows that $\pi$ provides a diffeomorphism from $N$ to
its image in $Q$; we need only see how $N$ being edgeless
makes this image all of $Q$.  Since $N$ is assumed embedded
and achronal, the easiest way to formulate ``edgeless" is
to take $N$ to be closed in $M$.

Since $\pi(N)$ is open, we need only prove it closed ($Q$
is connected, being the continuous image of the connected
space $M$).  Let $q$ be a point in $Q$, the limit of points
$\{\pi(x_n) = q_n\}$ with $x_n \in N$.  Pick a point $p$ in
$\gamma = \pi^{-1}(q)$, the Killing orbit corresponding to
$q$.  Let $P$ be a cross-section through $p$ of the
Killing-field foliation.  Let $\gamma_n$ be the foliate
corresponding to $q_n$; then each $\gamma_n$ intersects $P$
in precisely one  point $p_n$.  Let each $\gamma_n$ be
parametrized as an integral curve of the Killing field with
$\gamma_n(0) = p_n$, and  similarly parametrize $\gamma$
with $\gamma(0) = p$.  We then have  each $x_n =
\gamma_n(t_n)$ for some $t_n$.  Another way to formulate
that is in terms of the $R$-action on $M$ induced by the
Killing  action:  $x_n = t_n \cdot p_n$, where $s \cdot y$
denotes movement along the Killing orbit through $y$ by
parameter-value $s$.

All we need do now is show that the $\{t_n\}$ have an
accumulation point $t$, for then $x = \gamma(t) = t \cdot p$
will be a limit point of the ${x_n}$, so $x \in N$ and $q =
\pi(x) \in \pi(N)$.  But the only way for the $\{t_n\}$ to
avoid having an accumulation point is if they go off to plus
or minus infinity.  Consider any small neighborhood $V$
around $q$ in $Q$, and let $W$ be the tubular
neighborhood $\pi^{-1}(V)$ of $\gamma$ in $M$; we pick $V$
sufficiently small that $W$ is a standard static
spacetime.  For $x_n \in W$, the past and future null cones
from $x_n$ strike the side of $W$, so that the portion
$W_0$ of $W$ not timelike-related to $x_n$ is relatively
compact.  For all $m$ with $q_m \in V$, $x_m =
t_m \cdot p_m$ must lie in $W_0$ (since $N$ is
achronal).  But for $|s|$ sufficiently large,
$s \cdot p_m$ will lie outside $W_0$; thus, the $\{t_m\}$
must be bounded.
\qed\enddemo

Thus, in particular, when $M$ is globally hyperbolic, $Q$
has the topology of a Cauchy surface.  Note, however, that
when $M$ is static, it does not follow that the
restspaces---the hypersurfaces perpendicular to the Killing
field---have the topology of $Q$.  This is because the
restspaces need not, in general, be achronal (although any
restspace is necessarily a covering space of $Q$).  An
example would be the Minkowski cylinder $S^1\times \Bbb
L^1$ ($\Bbb L^n$ denotes Minkowski $n$-space), with Killing
field $U = d/dt +k(d/d\theta)$ for some small non-zero
constant $k$; the restspaces are spacelike helices
(topologically lines), while $Q$ is a circle.

The stationary observer space $Q$ comes equipped with a
natural Riemannian metric $h$, as shown by Geroch
\cite{5}:  The spacetime metric $g$ can be represented
(locally, in general; globally, if $Q$ is a Hausdorff
manifold) as $g = -\,(\Omega\circ\pi)\alpha^2 + \pi^*h$,
where $\Omega$ is a function on $Q$, $\pi$ is projection to
$Q$, $\alpha$ is the one-form obeying $\alpha U = 1$ and
$Ker(\alpha) =  U^\perp$, and $h$ is a Riemannian metric on
$Q$; $|U|^2 = \Omega\circ\pi$.  The spacetime is static if
and only if $d\alpha = 0$.

When $M$ is static, $h$ is the metric coming from the
projection of the restspaces to $Q$.  It turns out,
however, that this is not quite the appropriate metric to
use.  For many purposes, the appropriate metric on $Q$ is
$\bar h = \Omega^{-1}h$; call this the {\it conformal
metric\/} on the static observer space $Q$.  This is the
metric useful for calculating the causal properties of $M$
in terms of curves  on $Q$, and it also has this nice
property:

\proclaim{Theorem 4}
Let $M$ be a timelike or null geodesically complete,
static, globally hyperbolic spacetime.  Then the static
observer space $Q$ is complete in the conformal metric.
\endproclaim

\demo{Proof} We first examine the case in which $M$ is
simply connected:  The spacetime metric is $g =
(\Omega\circ\pi)\left[-\alpha^2 + \pi^*\bar h\right]$.
Since $(M,g)$ is globally hyperbolic it follows that
$(M,\bar g)$ is globally hyperbolic where $\bar g =
-\alpha^2 + \pi^*\bar h$.  Since $g$ is static, $d\alpha
= 0$.  Since $M$ is simply connected it follows that there
is a globally defined scalar function $\tau$ on $M$ such
that $d\tau = \alpha$.  The function $\tau$ measures
parameter-value along the integral curves of the Killing
field $U$; since this is complete (Lemma 1), the map $\tau
: M \to \Bbb R$ is onto.  Then $\pi$ and $\tau$ provide the
global identification $(M,\bar g) \cong (Q,\bar h)
\times \Bbb L^1$.  It then follows from proposition 2.54 of
reference \cite{8} that $(Q,\bar h)$ is complete.

We next consider the general case:  Let $\tilde M$ be the
universal covering space for $M$, with ${p_{_M}} : \tilde M
\to M$ the canonical projection.  We need to know that we
can apply the previous paragraph to $(\tilde M,
\tilde g)$, where $\tilde g = {p_{_M} ^*} g$.  A lemma to
this effect is in order.

\proclaim{Lemma 4.1}
Let $M$ be a spacetime and $\tilde M$ its universal
covering space, with metric induced from $M$.  Then $\tilde
M$ inherits all these properties from $M$:

1) static or stationary\par 2) energy conditions\par 3)
geodesic completeness of any type\par 4) global
hyperbolicity.
\endproclaim

\demo{Proof of 4.1} Any vector field on $M$ induces a
corresponding field on $\tilde M$; if the first is
Killing, so is the second, and the same goes for the
integrability of the perpendicular distributions.  Any
energy conditions are clearly inherited (the projection
being a local isometry), as is geodesic completeness of any
type (any appropriate geodesic segment projects to an
extendible segment, and the extension lifts to an
extension).

Suppose $M$ is globally hyperbolic.
There is a Cauchy surface $\Sigma$ in $M$.
Let $\tilde\Sigma = p^{-1}(\Sigma)$,
where $p: \tilde M \to M$ is the standard projection.
Then $\tilde\Sigma$ is a
Cauchy surface for $\tilde M$:  For
any inextendible timelike curve $\tilde c$ in $\tilde M$, $c =
p\circ\tilde c$ is an inextendible timelike curve in $M$.  For any
parameter-value $t$, $c(t) \in \Sigma$ if and only if
$\tilde c(t) \in \tilde\Sigma$.  Thus, $\tilde c$ intersects
$\tilde\Sigma$ in exactly one point, since that is true
for $c$ and $\Sigma$.  This shows that $\tilde M$ is also
globally hyperbolic.
\qed\enddemo

(Although not needed for our purposes here, it is perhaps
worth noting that the causality, chronology, and strong
causality properties are also inherited by universal
covering spaces.)
\smallpagebreak

By Lemma 4.1, we can apply the results of the first
paragraph to $(\tilde M, \tilde g)$:  Let $\tilde Q$ be
the space of static observers in $\tilde M$, with
$\tilde\pi: \tilde M \to \tilde Q$ the projection; then
$\tilde Q$ is complete in the conformal metric
$\bar{\tilde h}$ (notation here is that $\tilde g =
-\,(\tilde\Omega\circ\tilde\pi)\tilde\alpha^2 +
\tilde\pi^*\tilde h =
(\tilde\Omega\circ\tilde\pi)\left[-\tilde\alpha^2 +
\tilde\pi^*\bar{\tilde h}\right]$).  All we need to do is
compare this with $Q$ and its conformal metric $\bar h$.

First note that ${p_{_M}}$ induces a map ${p_{_Q}}: \tilde Q
\to Q$, since if $\tilde x$ and $\tilde y$ in $\tilde M$
lie on the same Killing orbit $\tilde\gamma$, then $x$ and
$y$ lie on the same Killing orbit $\gamma$, where $x =
{p_{_M}}\tilde x$, $y = {p_{_M}}\tilde y$, and $\gamma =
{p_{_M}}\circ\tilde\gamma$; we have ${p_{_Q}}\circ\tilde\pi =
\pi\circ {p_{_M}}$.  Also note that the metric induced on
$\tilde Q$ via ${p_{_Q}}$ from that on $Q$ is the same as
that which $\tilde Q$ inherits from the static structure
$\tilde\pi : \tilde M \to \tilde Q$ (because ${p_{_M}}$ is a
local isometry)---${p_{_Q} ^*} h = \tilde h$.  This applies
also to the respective conformal metrics (since the
conformal factor is the same in both
cases)---${p_{_Q} ^*} \bar h = \bar{\tilde h}$.  Thus, if we
can show that ${p_{_Q}}: \tilde Q \to Q$ is the standard
projection (universal covering map) of the universal
covering space for $Q$, then the completeness of $(\tilde
Q, \bar{\tilde h})$ will imply, by standard covering space
arguments, the completeness of $(Q, \bar h)$. Thus, we need

\proclaim{Lemma 4.2}
Let $M$ be a chronological stationary spacetime with a
complete Killing field; let $\pi: M \to Q$ be the
projection to the stationary orbit space.  Let ${p_{_M}}:
\tilde M \to M$ be the universal covering map from the
universal covering space for $M$, and let $\tilde\pi:
\tilde M \to \tilde Q$ be the projection to the stationary
orbit space.  Let ${p_{_Q}}: \tilde Q \to Q$ be the map
induced by ${p_{_M}}$ (i.e.,  ${p_{_Q}}\circ\tilde\pi =
\pi\circ {p_{_M}}$).  Then ${p_{_Q}}$ is the universal
covering map from the universal covering space for $Q$.
\endproclaim

\demo{Proof of 4.2}  By the results in \cite{6}, $Q$ is
a (Hausdorff) manifold; since $\tilde M$ inherits
chronology, stationarity, and completeness of the Killing
field from $M$, $\tilde Q$ is also a manifold.  The map
${p_{_Q}}$ is onto (an orbit in $M$ is mapped onto by the
corresponding orbit in $\tilde M$) and is locally a
homeomorphism (since ${p_{_M}}$ is, and the orbit spaces are
formed analogously in the two spacetimes).  Therefore, to
show that ${p_{_Q}}$ is a covering map, all that we need is
that it evenly covers small sets in $Q$, i.e., that for
any $q \in Q$, there is a neighborhood $U$ of $q$ such
that ${p_{_Q} ^{-1}} (U)$ is the disjoint union of sets
$\{V_\alpha\}$ in $\tilde Q$ with each restriction
${p_{_Q}}\bigr|_{V_\alpha}: V_\alpha \to U$ being a
homeomorphism.

Given $q \in Q$, pick any simply connected neighborhood $U$
of $q$ such that $\pi: M \to Q$ is locally trivial over
$U$, i.e., $\pi^{-1}(U) \cong U \times \Bbb R$.   Then
$\pi^{-1}(U)$ is simply connected also, so it is evenly
covered by ${p_{_M}}$ (a universal covering map evenly covers
an open set if and only if it is simply connected).  Thus,
${p_{_M} ^{-1}} (\pi^{-1}(U))$ is a disjoint union of open sets
$\{W_\alpha\}$ in $\tilde M$, each one carried
homeomorphically to $\pi^{-1}(U)$ by ${p_{_M}}$.  Let
$V_\alpha = \tilde\pi(W_\alpha)$ in $\tilde Q$.

In order to show each $V_\alpha$ is open, we need to show
$W_\alpha$ is full, i.e., contains only entire equivalence
classes of points; that means that $W_\alpha$ must contain
the entire orbit of any point it contains.  But if
$\tilde\gamma$ is an orbit in $\tilde M$ intersecting
$W_\alpha$, then $\gamma = {p_{_M}} \circ \tilde\gamma$ is an
orbit in $M$ intersecting $\pi^{-1}(U)$.  It follows that
$\gamma$ wholly lies in $\pi^{-1}(U)$, so $\tilde\gamma$
lies in the disjoint union of the $\{W_\beta\}$; since
these are disjoint open sets, $\tilde\gamma$ must lie
entirely in $W_\alpha$.  Thus, $V_\alpha$ is open.  Also,
the collection $\{V_\beta\}$ is disjoint:   $\{W_\beta\}$
is disjoint and made up of full sets.

It remains to be shown that, restricted to any $V_\alpha$,
${p_{_Q}}$ is a homeomorphism to $U$; this is straight-forward:
First, ${p_{_Q}}\bigr|_{V_\alpha}$ is onto $U$, because
${p_{_M}}(W_\alpha) = \pi^{-1}(U)$ and $\pi(\pi^{-1}(U)) = U$.
Next, it is injective:  If ${p_{_Q}}(\tilde q_1) =
{p_{_Q}}(\tilde q_2)$, then for any points $\tilde x_i$ in the
corresponding orbits $\tilde\gamma_i$ in $\tilde M$,
$\pi({p_{_M}}(\tilde x_1)) = \pi({p_{_M}}(\tilde x_2))$.
Since $\pi$ is trivial over $U$, this means that
${p_{_M}}(\tilde x_1)$ and ${p_{_M}}(\tilde x_2)$ lie on the
same orbit in $M$; since ${p_{_M}}$ is injective on
$W_\alpha$, that means the orbits $\tilde\gamma_1$ and
$\tilde\gamma_2$ must coincide, so
$\tilde q_1 = \tilde q_2$.  That ${p_{_Q}}$ is continuous
follows automatically from the commuting diagram
${p_{_Q}}\circ\tilde\pi = \pi\circ {p_{_M}}$.  Finally,
${p_{_Q}}$ is open:  An open set in $\tilde Q$ is precisely
$\tilde\pi(Z)$, where $Z$ is a full open set in $\tilde
M$; then ${p_{_Q}}(\tilde\pi(Z)) = \pi({p_{_M}}(Z))$.   Since
${p_{_M}}$ is an open map (being a covering projection),
${p_{_M}}(Z)$ is open in $M$.  Since $Z$ is made up of $\tilde
M$-orbits,
${p_{_M}}(Z)$ is made up of $M$-orbits, i.e., it is full;
therefore, $\pi({p_{_M}}(Z))$ is open in $Q$.

Thus, ${p_{_Q}}$ is a covering projection.  Since $\tilde M$ is
simply connected and $\tilde M \cong \tilde Q \times \Bbb
R$, it follows that $\tilde Q$ is simply connected:  ${p_{_Q}}$
is the universal covering map from the universal covering
space for $Q$.
\qed\enddemo

This completes the proof of Theorem 4.
\qed\enddemo

It is the complete Riemannian manifold $(Q,\bar h)$ that
provides the proper context to discuss the geometry of $M$.
\head
3. Results
\endhead
Let us now apply the energy condition.  For a static
spacetime the Killing vector $U$ is an eigenvector of
${R^a}_b$ (the linear map corresponding to $Ric$) (see
\cite{11}).  Call the corresponding eigenvalue $-\,S$, i.e., $S
= Ric(U,U)/|U|^2$.  Then we have

\proclaim{Proposition 5}
In a static spacetime obeying the null energy condition,
for any unit timelike vector $T$, $Ric(T,T) \ge S$.
\endproclaim

\demo{Proof} Let $V = U/|U|$, the unit vector in the
direction of $U$.  Let $X$ be any unit spacelike vector
orthogonal to $V$; then $V + X$ is a null vector.  The
null energy condition then yields $S + Ric(X,X) \ge 0$.

Now let $T$ be any unit timelike vector.  Then $T$ has the
form $T = aV + bX$ where $X$ is a unit spacelike vector
orthogonal to $V$ and $a$ and $b$  satisfy $a^2 - b^2
= 1$.  We then find
$$\align
Ric(T,T) &= a^2 S + b^2 Ric(X,X)\\
&= \left(a^2 - b^2\right)S + b^2\left[S + Ric(X,X)\right]\\
&\ge S.
\endalign$$
\qed \enddemo

For any point $q_0$ in $Q$ and any $r > 0$, let
$B_r(q_0) = \{q \in Q\,|\,d(q,q_0) \le r\}$, where $d$ is
the distance function from the conformal metric $\bar h$.
Note that since $(Q,\bar h)$ is complete, $B_r(q_0)$ is always compact.
Let $S_r(q_0) = \min_{B_r(q_0)}S$ ($S$ can
be thought of as a function on $Q$).  The main result of
this paper is

\proclaim {Theorem 6}
Let ($M,g$) be a static, globally hyperbolic, timelike
or null geodesically complete spacetime satisfying the
null energy condition.  Then for each point ${q_0} \in Q$,
for all $r > 0$, $S_r(q_0) \le K/r^2$,  where $K =
3\pi^2/\left(4|U|_{q_0}^2\right)$.
\endproclaim

\demo{Proof} Again, we first treat the case where $M$ is
simply connected. Given ${q_0} \in Q$, let $\gamma$
be the integral curve of $U$ corresponding to $q_0$.  For
each $r > 0$ let $\zeta$ be a unit-speed maximal geodesic
joining $\gamma(0)$ to $\gamma(2r)$.  The curve
$\zeta$ is timelike in the metric $\bar g = -d\tau^2 +
\pi^*\bar h$ (with notation the same as in the proof of
Theorem 4).  Since $\tau$ measures the parameter along
$\gamma$, it then follows that the projection of
$\zeta$ to $Q$ must remain in $B_r(q_0)$.  Going back
to the spacetime metric $g$ and using the result of
Proposition 5, we then find $Ric(\dot\zeta,\dot\zeta) \ge
S_r(q_0)$.  Since $\zeta$ is maximal it has no conjugate
points, and it has length at least $2r|U|_{q_0}$.  Then
applying the Lorentzian analogue of Myers' theorem
\cite{4} we find that $S_r(q_0) \le
3\pi^2/\left(4{|U|_{q_0}^2}\,r^2\right)$.

Now we consider the case where $M$ is not simply
connected.  As before, let $\tilde M$ be the universal
covering space of $M$ with projection ${p_{_M}}: \tilde M \to
M$ and $\tilde g$ the induced metric.  Then, by Lemma
4.1, $(\tilde M, \tilde g)$ is static, globally
hyperbolic, and timelike or null geodesically complete,
and satisfies the null energy condition.  With $\tilde
B_r$ and $\tilde S_r$ defined on the static observer space
$\tilde Q$, we have $\tilde S_r(\tilde q) \le
3\pi^2/\left(4|\tilde U|_{\tilde q}^2 \, r^2\right)$ for
any $\tilde q \in \tilde Q$.

By Lemma 4.2, the induced map ${p_{_Q}}: \tilde Q \to Q$
is a covering projection, and ${p_{_Q}}^*\bar h =
\bar{\tilde h}$.  It follows that for ${p_{_Q}}(\tilde q)
= q$ and ${p_{_Q}}(\tilde q_0) = q_0$, $\tilde d(\tilde
q,\tilde q_0) \ge d(q,q_0)$, where $\tilde d$ and $d$ are
the respective conformal distance functions (since
projection by $p_{_Q}$ preserves conformal lengths of
curves).  For any $q_0 \in Q$, pick a $\tilde q_0 \in
{p_{_Q}}^{-1}(q_0)$; then for any $r > 0$, there will be some
point $\tilde q \in \tilde B_ r(\tilde q_0)$ with
$\tilde S(\tilde q) \le K/r^2$ (where $K =
3\pi^2/\left(4|\tilde U|_{\tilde q_0}^2\right)$; since
$|\tilde U|_{\tilde q_0} = |U|_{q_0}$, this is the correct $K$).
Let $q = {p_{_Q}}(\tilde q)$; then $S(q) = \tilde S(\tilde q)$, and
$d(q,q_o) \le r$.  Therefore, $S_r(q_o) \le K/r^2$.
\qed \enddemo

Since $Q$ is complete in the conformal metric, we can
always find a direction---that is to say, a geodesic
emanating from $q_0$---such that, along that direction,
$S$ actually goes down at least as fast as $K/r^2$.

None of what we have done presupposes that $S$ is
actually positive, though the conclusion is vacuous if
$S$ is anywhere non-positive, once $r$ is large enough
for $B_r$ to contain any such point.  If we assume
positivity of $S$, we obtain a restriction on the spacelike
topology:

\proclaim{Corollary 7}
Let M be a static, globally hyperbolic, timelike or null
geodesically complete spacetime satisfying the
energy condition that $Ric(T,T) > 0$ for all timelike
vectors $T$.  Then the static observer space $Q$ is not
compact.
\endproclaim

\demo{Proof} This energy condition implies the null
energy condition, so Theorem 6 yields $S_r \le K/r^2$
for all $r$; this implies $\inf_Q S \le 0$.  On the
other hand, $S = Ric(U,U)/|U|^2$ is positive at all
points of $Q$, so if $Q$ were compact, $S$ would have to
have a positive infimum.
\qed\enddemo

\head
4. Discussion
\endhead

We expect some result along the lines that in order to avoid
gravitational collapse, spacetime must not contain ``too
much matter.'' This naive expectation is made more precise by
noting that global hyperbolicity and geodesic completeness put
constraints on the Ricci tensor, specifically, on $Ric(T,T)$
where $T$ is the unit tangent vector to a maximal timelike
geodesic segment:  This component of $Ric$ is obliged,
somewhere along any such geodesic segment, to be less than
an inverse-quadratic function of the length of the segment.

The specific form of the results we have obtained depends upon
the spacetime being foliated by a system of ``canonical''
oberservers, with measurement being made with respect to this
canonical observer space.  For any such foliated spacetime, if
there are sufficient completeness conditions to assure that
there is a maximal geodesic between any pair of points on each
observer orbit, and that the observer orbits are infinitely
long, then we will be able to conclude that there will be
points of small Ricci-value associated with each observer.
The difficulty comes in finding a geometric framework for
these points.  In physical terms, what is wanted is this:
Given two events on the worldline of a base-point canonical
observer, consider an arbitrary observer moving from one event
to the other one; we need to be able to say how far that
observer can wander in the canonical observer space.  If we
can place a bound on that wandering, in terms of the proper
time between those two events, then we can say that the place
of small Ricci-value associated to the maximal geodesic
between those events occurs (in the canonical observer space)
within that bound from the base-point.

In this paper we have considered static spacetimes, largely
because we can then easily bound the wandering of an arbitary
observer between events on the worldline of a static
observer.  Even a generalization to stationary spacetimes
becomes problematical due to the difficulty of finding such a
bound.

{}From the conditions used in theorem 6 one might have hoped
for a stronger bound on the behavior of the Ricci tensor.
We have placed bounds on only one component of $Ric$.
Furthermore we have shown only that that component must
fall off in one direction rather than in all directions.
We now consider two examples that show that stronger bounds of
this sort do not apply. First consider the Einstein
Static Universe \cite{1} whose metric is given by
$g = - \,d t^2\, + \, a^2h_{S^3}$, where $a$ is a constant and
$h_{S^3}$ is the metric of the unit three-sphere.  This
spacetime is a homogeneous, isotropic, and static
universe with Killing field $U = d/dt$. The matter in this
universe is dust and a cosmological constant.  The
Einstein Static Universe satisfies all the hypotheses of
Theorem 6; it also satisfies $Ric(U,U) = 0$, i.e., $S =
0$.  However, $Ric(X,X) = 2/{a^2}$ for any unit vector $X$
orthogonal to $U$. Thus $S$ vanishes, but the other
components of $Ric$ do not fall off at all.

Our second example is Melvin's Magnetic Universe \cite{9}
whose  metric is given by
$$d s^2 = F^2\left(- \,d t^2\, + \,d\rho^2\, + \,dz^2\right)
\; + \; F^{-2}\rho^2 \,d\phi^2 \; .$$
Here $F = 1 + {{( {b_0} \rho /2)}^2}$ where $b_0$ is
a constant. This spacetime is a static, cylindrically
symmetric solution of the Einstein-Maxwell equations.  It
represents an infinitely long tube of magnetic flux held
together by its own gravity.  The constant $b_0$ is the
value of the magnetic field on the axis.  Melvin's Magnetic
Universe satisfies all of the hypotheses of Theorem 6.
The static observer space with the conformal metric is
given by $(Q,\bar h) = (\Bbb R^3, d\rho^2\, + \,dz^2\, +
\,F^{-4}\rho^2\, d\phi^2)$.  Straightforward calculation shows
that $S = b_0^2/F^4$ and that, with $q_0$ given by $\rho = z = 0$,
$S_r = b_0^2/\left[1\, + \,{({b_0}r/2)}^2\right]^4$.
We see that $S_r$ falls off at an appropriate rate in $r$,
but that $S$ falls off only as $\rho$ goes to infinity and
is independent of position along the cylindrical axis (the
$z$ direction).  Thus spacetimes satisfying the hypotheses
of Theorem 6 need not have $S$ falling off in all
directions.
\head
Acknowledgements
\endhead
We would like to thank Gary Horowitz and Mael Melvin for
helpful discussions. We would also like to thank the Aspen
Center for Physics for hospitality. This research was
supported in part by NSF grant  DMS9310477 to St. Louis
University, NSF Grant PHY9408439 to Oakland University and
by a Cottrell College Science Award of Research Corporation
to Oakland University.
\enddocument